\documentclass[]{article}

\usepackage{graphicx}
\usepackage{amsmath}
\usepackage{physics}
\usepackage{authblk}
\usepackage[labelfont=bf]{caption}
\usepackage{hyperref}
\usepackage[margin=1in]{geometry}
\usepackage[superscript,biblabel]{cite}

\newcommand{\beginsupplement}{
        \setcounter{figure}{0}
        \renewcommand{\thefigure}{S\arabic{figure}}
     }

\title{Quantum Interference between Transverse Spatial Waveguide Modes}
\author[1,2]{Aseema Mohanty}
\author[1,3]{Mian Zhang}
\author[1,2]{Avik Dutt}
\author[4,5]{Sven Ramelow}
\author[6]{Paulo Nussenzveig}
\author[1,2,7,*]{Michal Lipson}
\affil[1]{School of Electrical and Computer Engineering, Cornell University, Ithaca, New York 14853, USA}
\affil[2]{Department of Electrical Engineering, Columbia University, New York, New York 10027, USA}
\affil[3]{School of Engineering and Applied Sciences, Harvard University, Cambridge, Massachusetts 02138, USA}
\affil[4]{School of Applied and Engineering Physics, Cornell University, Ithaca, New York 14853, USA}
\affil[5]{Faculty of Physics, University of Vienna, 1090 Vienna, Austria}
\affil[6]{Instituto de Fisica, Universidade de S\~ao Paulo, P.O. Box 66318, S\~ao Paulo 05315-970, Brazil}
\affil[7]{Kavli Institute at Cornell for Nanoscale Science, Cornell University, Ithaca, New York 14853, USA}
\affil[*]{Corresponding Author: ml3745@columbia.edu}

\date{\vspace{-5ex}}

\begin{document}

\maketitle

\section{Introduction}

\textbf{Integrated quantum optics has drastically reduced the size of table-top optical experiments to the chip-scale, allowing for demonstrations of large-scale quantum information processing and quantum simulation\cite{politi_silica--silicon_2008,metcalf_quantum_2014,spring_boson_2013,crespi_integrated_2013,broome_photonic_2013,tillmann_experimental_2013,aaronson_computational_2011}.  However, despite these advances, practical implementations of quantum photonic circuits remain limited because they consist of large networks of waveguide interferometers that path encode information which do not easily scale. Increasing the dimensionality of current quantum systems using higher degrees of freedom such as transverse spatial field distribution, polarization, time, and frequency to encode more information per carrier will enable scalability by simplifying quantum computational architectures\cite{lanyon_simplifying_2009}, increasing security and noise tolerance in quantum communication channels\cite{groblacher_experimental_2006,zhong_photon-efficient_2015}, and simulating richer quantum phenomena\cite{neeley_emulation_2009}. These degrees of freedom have previously been explored in free-space and fiber quantum systems to encode qudits and implement higher dimensional entanglement\cite{mair_entanglement_2001,langford_measuring_2004,barreiro_generation_2005,nagali_optimal_2009,karimi_exploring_2014}. Here we demonstrate a scalable platform for photonic quantum information processing using waveguide quantum circuit building blocks based on the transverse spatial mode degree of freedom: mode multiplexers and mode beamsplitters.  A multimode waveguide is inherently a densely packed system of spatial and polarization modes that can be coupled by perturbations to the waveguide. We design a multimode waveguide consisting of three spatial modes (per polarization) and a nanoscale grating beamsplitter to show tunable quantum interference between pairs of photons in different transverse spatial modes. We also cascade these structures and demonstrate NOON state interferometry within a multimode waveguide. These devices have potential to perform transformations on more modes and be integrated with existing architectures, providing a scalable path to higher-dimensional Hilbert spaces and entanglement.}

Currently, integrated quantum photonic circuits are primarily limited to path encoding information, but the use of a higher-dimensional Hilbert space within each path will increase the information capacity and security of quantum systems. Higher dimensionality allows one to encode more information per photon, relieving resource requirements on photon generation and detection.\cite{groblacher_experimental_2006,zhong_photon-efficient_2015} Consequently, this leads to more efficient logic gates and noise resilient communications, making quantum systems more scalable and practical.\cite{gisin_quantum_2002,lanyon_simplifying_2009} In integrated schemes, a few demonstrations have been developed for polarization \cite{corrielli_rotated_2014} and time\cite{xiong_compact_2015}. In free-space optics, orbital angular momentum and Hermite-Gaussian modes have both been used to encode information within a higher-dimensional space as qudits (d-level logic units)\cite{mair_entanglement_2001,langford_measuring_2004,barreiro_generation_2005,nagali_optimal_2009,karimi_exploring_2014}. However, the great potential of higher-order waveguide modes to encode quantum information has not been demonstrated to date\cite{saleh_modal_2009,bharadwaj_generation_2015}. The transverse spatial degree of freedom is an untapped resource that can be manipulated using simple photonic structures and does not require exotic material properties. 

\begin{figure}[htbp]
\includegraphics[width=\textwidth]{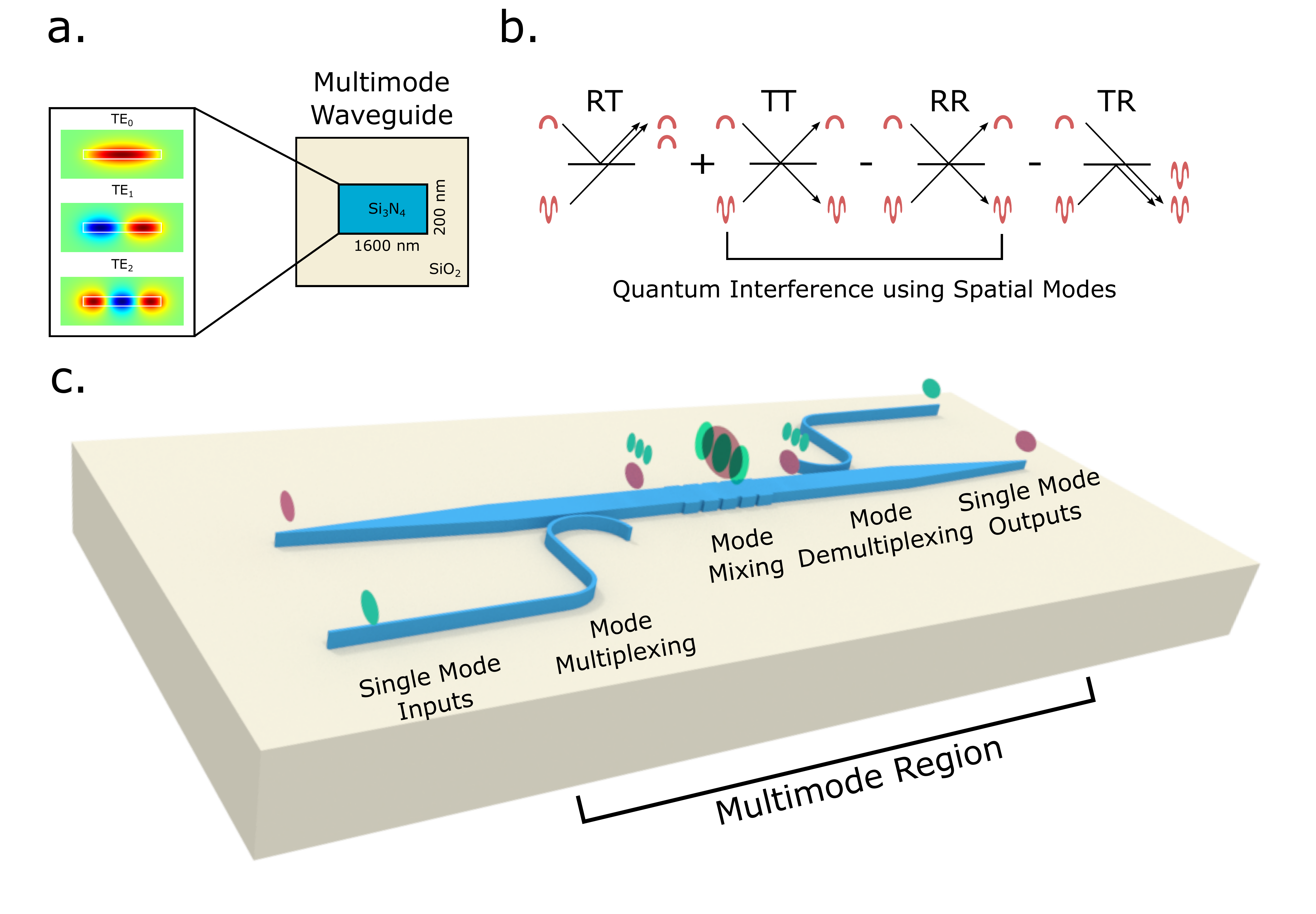}
\centering
\caption{\textbf{Quantum Interference using a Spatial Mode Beamsplitter} a) Simulation of transverse spatial modes in a multimode waveguide.  b) Concept of interference in a beamsplitter for indistinguishable paths using spatial modes c) Schematic showing chip implementation of mode multiplexing (asymmetric directional coupler) and mode beamsplitter (nanoscale grating).  The colors indicate the path between single mode inputs or outputs to different spatial modes in the multimode waveguide. Wavelength and polarization are identical within each path.}
\label{fig:1concept}
\end{figure}

In the classical regime, the orthogonal spatial modes of an integrated waveguide have been shown to dramatically scale data transmission rates\cite{luo_wdm-compatible_2014,wang_improved_2014,dorin_two-mode_2014,heinrich_supersymmetric_2014,stern_-chip_2015}. A waveguide can support many co-propagating modes, which can be used as parallel channels within a single waveguide. As an example, we consider a silicon nitride multimode waveguide with a sub-micron cross-section containing six modes: three spatial modes per polarization (see Figure \ref{fig:1concept}a). Progress in the field has overcome the challenge of achieving controlled coupling while avoiding unwanted coupling between different modes, for example in bends\cite{gabrielli_-chip_2012} and tapers\cite{dai_mode_2012}. Mode conversion based on waveguide structuring has significant potential in the quantum regime.\cite{ohana_mode_2014,tseng_implementation_2006,lu_objective-first_2012}.

\section{Experiment}

In order to show the potential utility of the integrated transverse spatial degree of freedom for scalable quantum information processing, we demonstrate Hong-Ou-Mandel (HOM) interference between two different quasi-transverse electric (TE) waveguide modes (TE0 and TE2). HOM interference is a useful proof of principle because it is the basis of many other quantum operations such as higher-dimensional entanglement, teleportation, quantum logic gates, and boson-sampling\cite{politi_silica--silicon_2008,metcalf_quantum_2014,spring_boson_2013,crespi_integrated_2013, nagali_experimental_2010,nagali_optimal_2009,karimi_exploring_2014}. In the original HOM experiment, a path beamsplitter is used to combine two originally orthogonal paths of two single photons, making them indistinguishable. The probability amplitudes of the two cases that contribute to detection of the two photons in coincidence destructively interfere owing to the bosonic nature of photons, if the two paths are indistinguishable.\cite{hong_measurement_1987} In our experiment, we replace the path beamsplitter with a spatial mode beamsplitter and replace the two paths with two spatial modes within a multimode waveguide (see Figure \ref{fig:1concept}b). The spatial mode beamsplitter couples two different spatial modes, resulting in a superposition of the two spatial modes. Mode coupling leads to interference within the waveguide between the cases in which both photons remain in their original modes or both couple to opposite modes (cases RR and TT in Figure \ref{fig:1concept}b). Visibility of the interference in coincidences is a measure of the equal splitting in the beamsplitter and indistinguishability of the two paths in every degree of freedom including transverse spatial mode.

The key building blocks required to demonstrate HOM interference are 1) a mode multiplexer for generating the different spatial modes and 2) a mode beamsplitter for interfering the modes, which both rely on selective mode coupling by phase matching in our design. The mode-multiplexer allows us to generate orthogonal spatial modes within the multimode waveguide without cross-talk between the modes, which would reduce the interference visibility. We couple pairs of photons from a spontaneous parametric downconversion (SPDC) source into single-mode silicon nitride waveguides that couple into a multimode waveguide (see methods and Figure \ref{fig:1concept}c and Figure \ref{fig:1extsetup}). Finally, the photons are sent to the spatial mode beamsplitter where they are equally split between the two modes, coupled into single mode output waveguides, and detected as coincidences.  We use a silicon nitride platform because the high core-cladding (Si$_3$N$_4$/SiO$_2$) index contrast allows one to strongly vary the propagation constants of different spatial modes by varying the waveguide dimensions, which is essential for selective mode coupling. The silicon nitride platform is attractive for integrated quantum information processing because its transparency window spans the visible to mid-infrared wavelength range and has been used to demonstrate non-classical light sources\cite{ramelow_silicon-nitride_2015,dutt_-chip_2015}.

\begin{figure}[htbp]
\includegraphics[width=\textwidth]{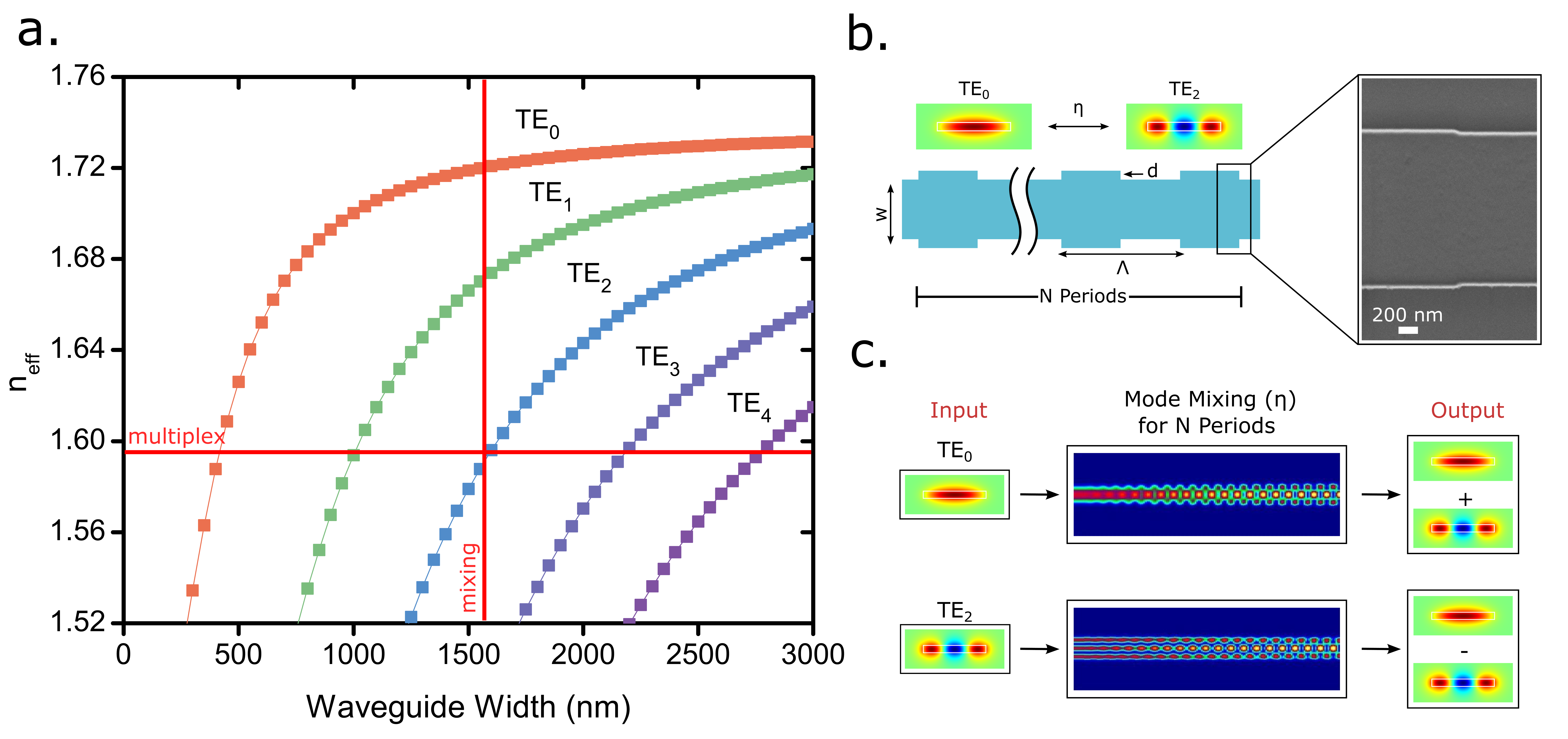}
\centering
\caption{\textbf{Design of Spatial Mode Beamsplitter} a) Si$_3$N$_4$ dispersion for a multimode waveguide with 190 nm height. Horizontal red line shows phase matching for waveguides with different widths for mode multiplexing. Vertical red line shows phase matching between modes in a single waveguide for the mode beamsplitter. b) Symmetric grating structure for coupling the TE0 and TE2 modes. The period is defined by the difference in effective index between the modes in a particular waveguide. The period ($\Lambda$) is 6.675 $\mu m$, and grating depth, d, is 24 nm. The width, w, is 1600 nm and height is 190 nm. Inset: SEM of fabricated grating structure. c) Simulation of mode conversion in a 50:50 splitter, or $\eta = 0.5$, where N = 20 periods.}
\label{fig:2design}
\end{figure}

To demonstrate the mode multiplexer, we use an asymmetric directional coupler to selectively couple the fundamental mode in a single-mode waveguide to a specific higher-order mode in an adjacent multimode waveguide. The asymmetric directional coupler uses two different waveguide widths to phase match light propagating in different modes within adjacent waveguides, allowing for efficient coupling.\cite{luo_wdm-compatible_2014,wang_improved_2014} In Figure \ref{fig:2design}a, the horizontal red line indicates where the effective indices of different higher-order modes in waveguides of different widths match. For example, to excite the TE2 mode in the multimode waveguide using the TE0 mode in a single mode waveguide with 420 nm width, we choose the multimode waveguide width of 1.6 ${\rm\mu m}$ (see methods).

To demonstrate the mode beamsplitter, we use a nanoscale grating structure to selectively couple different higher-order spatial modes within a multimode waveguide. The period of the grating structure provides a momentum change that accounts for the phase mismatch between the two different spatial modes\cite{yariv_coupled-mode_1973}. In Figure \ref{fig:2design}a, the vertical red line indicates the phase mismatch ($\Delta n_{eff}$) between modes within a single waveguide, $ \Lambda = \lambda/\Delta n_{eff}$ where $\Lambda$ is the period of the grating, $\lambda$ is the wavelength, and $n_{eff}$ is the effective index of the mode. For example, to couple TE0 and TE2, $\Delta n_{eff} = 0.12$ and $\Lambda = 6.675 \mu m$. We define splitting ratio, $\eta$, as the probability of coupling to the same mode, and 1-$\eta$ as the probability of coupling to the opposite mode. This splitting ratio can be tuned from 0 to 100\% if the two modes are perfectly phase matched. This splitting ratio ($\eta$) depends on the coupling coefficient ($\kappa$) determined by the overlap of the two modes within the perturbed region (grating depth, d) and the length of coupling interaction (or the number of periods, N) as follows: $\eta = sin(\kappa N)^2$ (see Figure \ref{fig:2design}b). We use Finite Element Method and EigenMode Expansion to determine the phase matching and splitting ratios. Figure \ref{fig:2design}c shows a simulation of a 50:50 coupler between TE0 and TE2. Beamsplitters with tunable splitting ratio are crucial building blocks for photonic quantum simulation circuits\cite{ma_quantum_2011,spring_boson_2013,crespi_integrated_2013} and for reconfigurable quantum circuits for quantum metrology and processing.\cite{metcalf_quantum_2014} 

\section{Results}
\begin{figure}[htbp]
\includegraphics[width=\textwidth]{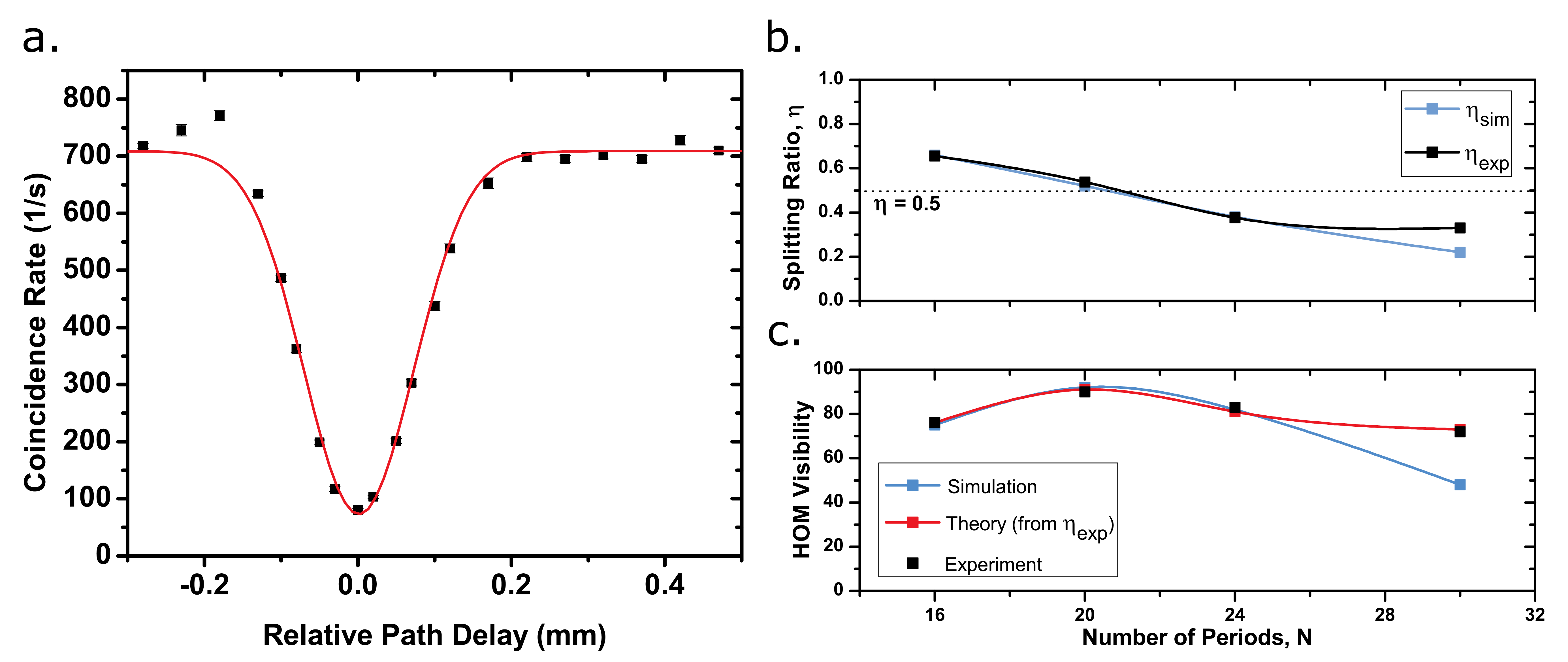}
\centering
\caption{\textbf{Hong-Ou-Mandel Interference between TE0 and TE2} a) We show the coincidence rate (accidentals subtracted) of the two output arms as we delay one input arm. The red line is a gaussian fit to the experimental data. The HOM visibility is $90 \pm 0.8$\%. The error bars come from standard error and are not visible because they are smaller than the data points. b) Experimental and simulation comparison of the splitting ratio as the number of periods (N) is varied.  Error bars on experimental data are smaller than data points. c) Corresponding HOM visibility as the number of periods (N) is varied.
}
\label{fig:3homresults}
\end{figure}

We observe a high HOM visibility of $90 \pm 0.8$\% between photons sent through the TE0 and TE2 mode channels. In Figure \ref{fig:3homresults}a, coincidences with accidentals subtracted are plotted against relative path length difference between the two input arms, and the best gaussian fit is indicated by the red curve. The device with splitting ratio near 1/2 (where N=20), yields the highest visibility of $90 \pm 0.8$\% with a coherence length of $168 \pm 10 \mu m$, which we estimate from the width of the coincidence dip. This device is primarily limited by the source visibility, which we measure to be 92\%(see methods) due to spectral mismatch of the two arms. With an ideal source, this device could have a high visibility of 99\% with a measured splitting ratio of $\eta_{exp}= 0.55$. In Figure \ref{fig:3homresults}b, we show measured splitting ratios near 1/3, 1/2, and 2/3 for devices with different numbers of coupling periods, which agree well with simulation. These ratios have been of particular importance in path-encoded implementations of controlled-NOT gates in quantum photonic circuits\cite{politi_silica--silicon_2008}. Note that the device with the longest coupling interaction does not produce as much splitting as predicted by simulation, which is most likely due to residual phase mismatch. As expected, we show that the experimental HOM visibilities depend on the splitting ratios measured and agree well with their theoretical visibilities from the measured splitting ratios (Figure \ref{fig:3homresults}c). To show that this method easily extends to other modes of different parities, we also demonstrate a visibility of $78 \pm 0.3$\% between TE0 and TE1. We use an asymmetric grating for a structure that is limited to 78\% by its splitting ratio ($\eta = 0.64$)(see extended Figure \ref{fig:2extsetup})

\begin{figure}[htbp]
\includegraphics[width=\textwidth]{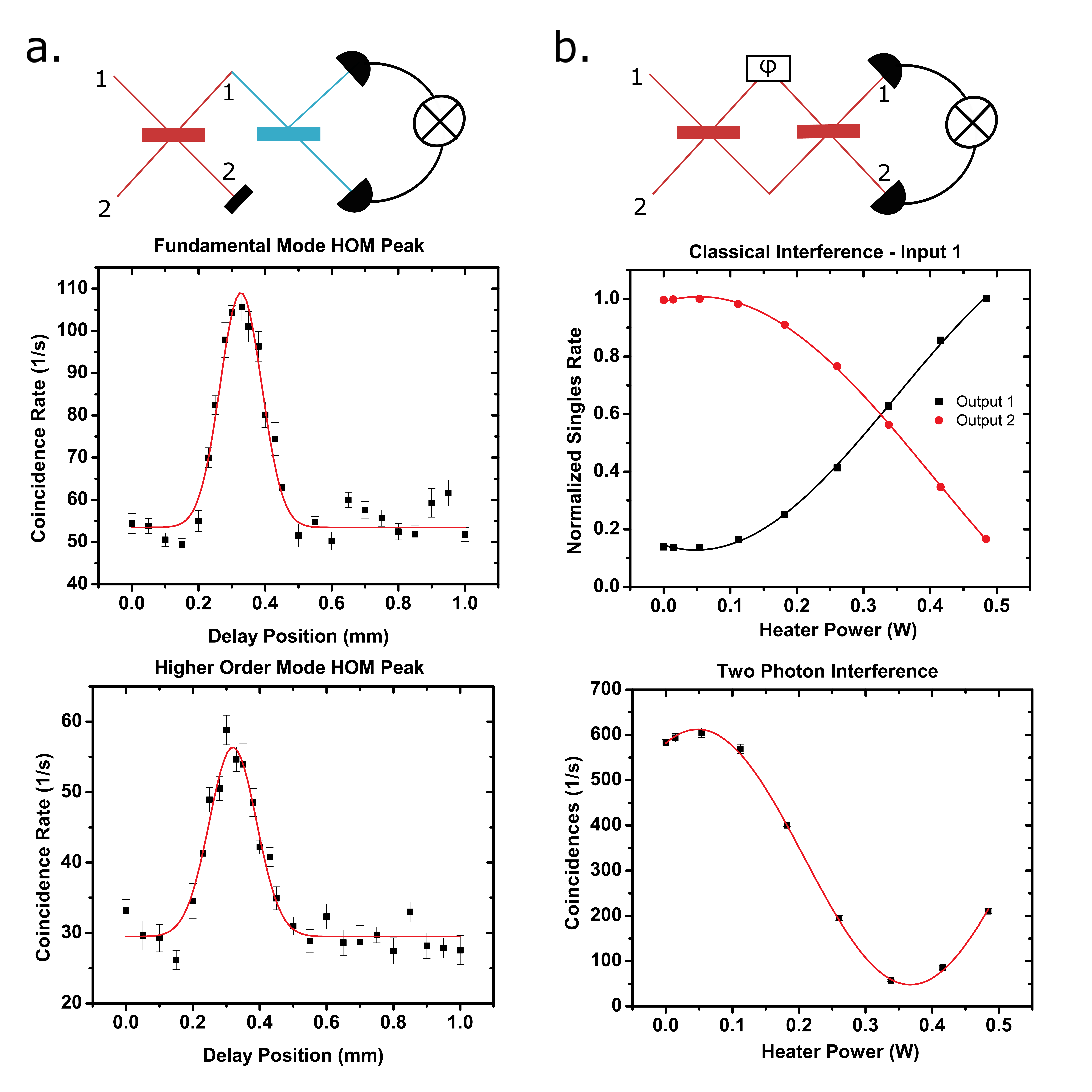}
\centering
\caption{\textbf{Hong-Ou-Mandel Peak and NOON Interference} a) Coincidence rate of individual HOM output arms after fiber beamsplitter. Red indicates on-chip spatial mode beamsplitter, and blue indicates fiber beamsplitter. There is a peak in coincidences due to HOM bunching at each output. b) Classical Mach-Zehnder fringe (top) and NOON interference (bottom) is shown as a function of the power applied to the integrated heater. The period of the quantum interference is half that of the classical interference.
}
\label{fig:4peakandnoon}
\end{figure}

\begin{figure}[htbp]
\includegraphics[width=\textwidth]{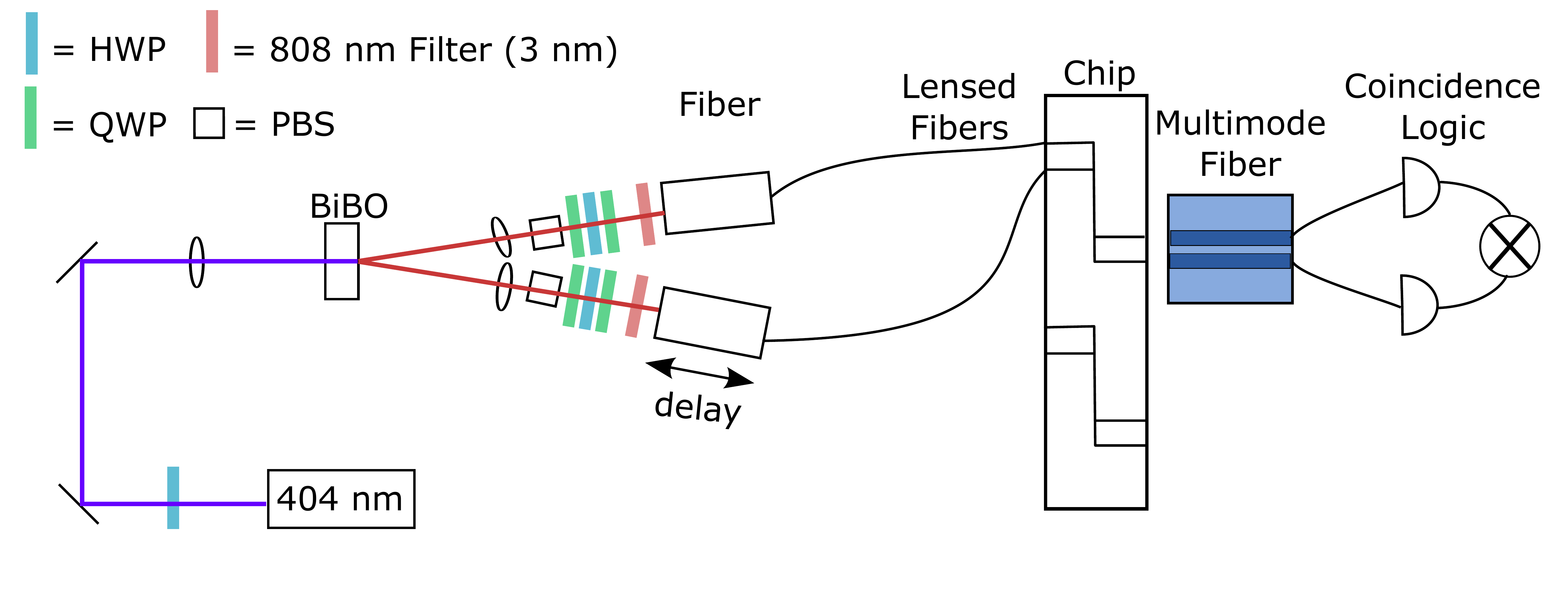}
\centering
\caption{\textbf{Spontaneous Parametric Down Conversion (SPDC) source and coincidence counting setup.}
}
\label{fig:1extsetup}
\end{figure}

\begin{figure}[htbp]
\includegraphics[width=\textwidth]{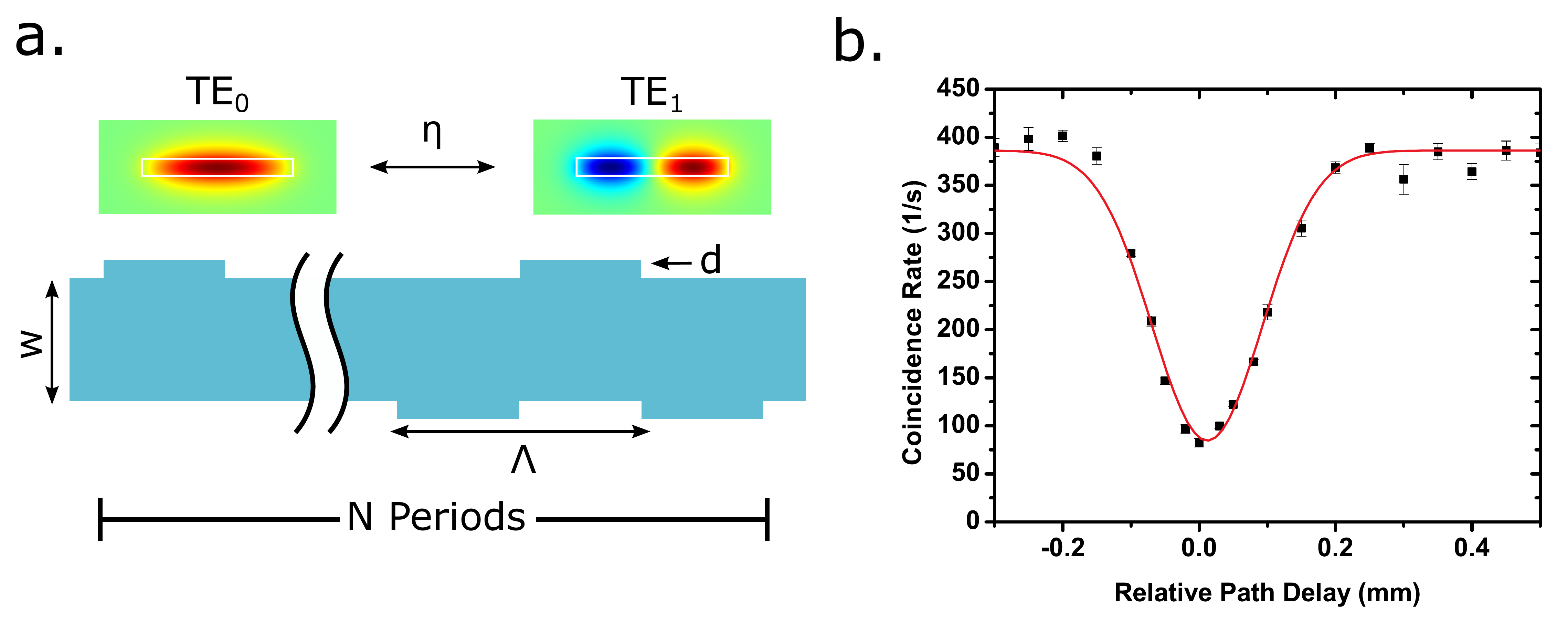}
\centering
\caption{\textbf{Hong-Ou-Mandel Interference between TE0 and TE1} a) Schematic of asymmetric grating to couple even to odd modes. b) We show the coincidence rate (accidentals subtracted) of the two output arms as we delay one input arm. The red line is a gaussian fit to the experimental data. The HOM visibility is $78 \pm 0.3$\%.
}
\label{fig:2extsetup}
\end{figure}

To further confirm the observed Hong-Ou-Mandel effect, we measure photon coalescence enhancement at the individual output arms of the HOM interferometer\cite{nagali_optimal_2009,karimi_exploring_2014}. We expect a doubling of the probability of case TR and RT in the HOM experiment in comparison to the experiment with distinguishable photons (see Figure \ref{fig:1concept}a). We use a fiber beamsplitter at the individual output arms of the spatial mode beamsplitter ($\eta = 0.55$) and measure coincidences. Figure \ref{fig:4peakandnoon}a shows a peak in coincidences for both the fundamental and higher order mode output port with a visibility of $2 \pm 0.02$ that matches well with theory. The width of the multimode HOM peak is $166 \pm 10 \mu m$, and the width of the single mode output is $147 \pm 10 \mu m$. This effect has been used as a basis for quantum cloning experiments\cite{nagali_optimal_2009}.

Finally, to show these structures can be cascaded and actively tuned, we fabricate a Mach-Zehnder structure to create a NOON state interferometer based on our spatial mode beamsplitter\cite{giovannetti_quantum-enhanced_2004}. The HOM interferometer and phase shifter produces the NOON state described by: $\frac{1}{\sqrt{2}}(\ket{2}_1\ket{0}_2+e^{2i\phi}\ket{0}_1\ket{2}_2)$ where $\phi$ is the phase between the two modes of the interferometer and the subscripts 1 and 2 refer to the different modes. NOON states are more generally written as $\frac{1}{\sqrt{2}}(\ket{n}_1\ket{0}_2+e^{in\phi}\ket{0}_1\ket{n}_2)$ and provide increased phase sensitivity, $\phi$, by $\frac{1}{n}$ for quantum metrology over the standard quantum limit of $\frac{1}{\sqrt{n}}$\cite{giovannetti_quantum-enhanced_2004}.  In our experiment, the Mach-Zehnder structure consists of two gratings separated by a phase shifter, a length of waveguide and heater (see methods). Within the phase-shifter, the waveguide is tapered out to 10 $\mu m$ width which gives a larger differential in phase shift between the fundamental and higher order modes as the heater is tuned. In Figure \ref{fig:4peakandnoon}b, we measure the classical interference by inputting a single arm of the SPDC source into the device and measuring the singles counts of both the output arms, which show the classical Mach-Zehnder fringe as expected. This specific device ($\eta_{exp}= 0.66$) has a classical visibility of $82 \pm 8$\% with a period of about  $1.3 \pm 0.082$ W, which corresponds to the power of the heater. We then measure the two-photon interference, or NOON state interference, by measuring coincidences when both arms of the SPDC source are input into the device with no path delay. We observe a visibility of $86 \pm 1$\% with a period of $0.64 \pm 0.005$ W, about half of the classical interference. In addition to the increased phase sensitivity, this demonstrates the active tunability of this device, which could be useful in state preparation for quantum simulators \cite{ma_quantum_2011,spring_boson_2013,crespi_integrated_2013}.

\section{Discussion}

In this paper, we show a step perturbation that has a frequency response that includes additional higher order frequencies. Because we have limited our multimode waveguides to the lowest three ordered modes, these higher frequency components do not pose problems. When dealing with a larger number of modes that require couplings given by multiple spatial frequency components, a sinusoidal perturbation would ensure less cross talk between the modes. These devices for two-mode couplings could be cascaded to create arbitrary transformations between modes\cite{zukowski_realizable_1997}. This initial demonstration between two modes can be extended to make arbitrary n-dimensional unitary matrix transformations on a set of modes, which is essential for quantum information processing and simulation\cite{zukowski_realizable_1997}. Assuming 5 nm fabrication tolerance on dimensions, we can realistically expect to multiplex at least 15 modes within a silicon nitride waveguide\cite{zortman_silicon_2010}. This number of quantum modes corresponds to a Hilbert space with a dimensionality of $15^2 = 225$ for a two waveguide system.  Higher index materials will increase the number of modes that can be multiplexed. These designs could also be made more compact by using multiplexed gratings in which the perturbation has multiple spatial frequency components and strengths to design arbitrary transformations of the modes\cite{ohana_mode_2014,tseng_implementation_2006,lu_objective-first_2012}.

We show that these structures are tunable and can be cascaded while preserving high visibility quantum interference, which will be key to building larger networks. Multimode waveguides can be used with the other degrees of freedom to encode information within a high-dimensional Hilbert space using only linear passive devices within a small footprint. These massively scaled systems could eventually be interfaced with mode multiplexing in fiber and free-space systems for quantum information processing and could be specifically useful in quantum repeaters, memories, and simulators.

\section{Methods}
\textbf{Device Design and Characterization:} 
\newline
We use silicon nitride waveguides to implement the mode multiplexing and spatial mode beamsplitter. The device has inverse tapers ($\sim$170 $\mu m $) to mode-match to 2 $\mu m $ spot size of tapered fibers. The single mode waveguides are 190 nm tall and 420 nm wide. The single mode waveguide is tapered adiabatically (100 $\mu m $ long taper) to the multimode waveguide, which is 190 nm tall and 1600 nm wide. We use COMSOL and FIMMWAVE software packages to simulate the mode profiles and coupling. The asymmetric directional coupler has a coupling length of 18 $\mu m $ between the single mode and multimode waveguide. The perturbation needed to couple the modes in the multimode waveguide is quite small, about 24 nm, in order to remain within the weak coupling regime (see fig. 2b), but large enough to yield reasonable device lengths. For gratings with 24 nm depth, the coupling coefficient is  ($\kappa = 0.041$) per period. The simulation shows approximately 50:50 coupling for N=20, corresponding to a device length of about 133 $\mu m$ for our specific geometry. We estimate the loss in the device excluding coupling losses to be 0.2 dB. To characterize the on-chip beamsplitters and fiber beamsplitters, the classical splitting ratios ($\eta$) were measured using an 808 nm diode laser source.
\newline
\newline
\textbf{Device Fabrication:} 
\newline
We deposit 190 nm of low-pressure chemical vapor silicon nitride on a silicon wafer with 4 $\mu m$ of thermal oxide. Then, we pattern with electron beam lithography and etch the waveguides. We finally clad the devices with 300 nm of high temperature oxide and 2 $\mu m$ of plasma enhanced chemical vapor deposited silicon dioxide. For the cascaded device with the integrated phase shifter, we fabricate the heater (50 nm Ni) and contact pads (200 nm Al) using a metal lift-off process.
\newline
\newline
\textbf{Experimental Setup for Interference:} 
\newline
To observe the HOM interference, we couple photon pairs into the spatial mode beamsplitter on the chip and measure coincidences. We produce degenerate 808 nm photon pairs using spontaneous parametric downconversion (SPDC, Type I) by pumping a Bismuth Borate (BiBO) crystal with a 404 nm diode laser (see extended figure 1). We use polarizing beamsplitters, waveplates, and bandpass filters (3 nm FWHM) to couple indistinguishable photon pairs into the chip using lensed fiber. The output of the beamsplitter is collected using a multimode fiber array and sent to the single photon counting modules (SPCM-AQRH) and coincidence logic (Roithner TTM8000). We manually adjust the delay by translating one of the fiber couplers at the source with a micrometer screw and use a coincidence window of 2 ns to minimize accidental coincidences. The SPDC source HOM visibility was characterized using a single mode fiber beamsplitter, and we measured a visibility of $92 \pm 1.9$\% and coherence width of $194 \pm 10 \mu m$. We attribute this reduced visibility primarily to spectral differences between the two arms. The HOM peak experiment used a multimode fiber beamsplitter and detected coincidences with the same coincidence window. Finally, to test the Mach-Zehnder structure, we apply a voltage on the heater using a Keithley sourcemeter to produce the phase shift between the modes.

\section{Acknowledgment}
This work was supported by the National Science Foundation - CIAN ERC (EEC-0812072). This work was performed in part at the Cornell NanoScale Facility, a member of the National Nanotechnology Coordinated Infrastructure (NNCI), which is supported by the National Science Foundation (Grant ECCS-15420819). P.N. acknowledges funding from Brazilian agencies CNPq and FAPESP. A.M. was funded by a National Science Foundation Graduate Research Fellowship (Grant No. DGE-1144153), and S.R. was funded by an EU Marie Curie Fellowship (PIOF-GA-2012-329851).

\newpage

\newpage
\section{Supplementary Figures}

\beginsupplement

\end{document}